\documentclass[12pt,a4paper]{article}
\usepackage[utf8]{inputenc}
\usepackage[T1]{fontenc}
\usepackage{times}
\usepackage[margin=1in]{geometry}
\usepackage{graphicx}
\usepackage{amsmath,amssymb}
\usepackage{booktabs}
\usepackage{caption}
\usepackage{subcaption}
\usepackage{float}
\usepackage{hyperref}
\usepackage[numbers]{natbib}
\usepackage{setspace}
\usepackage{enumitem}
\usepackage{multirow}
\usepackage{array}
\usepackage{tabularx}
\usepackage{titlesec}
\usepackage{soul}
\usepackage{xcolor}
\makeatletter
\renewcommand{\maketitle}{
\begin{center}
\colorbox{white}{%
  \begin{minipage}{0.92\textwidth}
    \centering
    \fontsize{12}{14}\selectfont\bfseries \@title
  \end{minipage}%
}\par
\vspace{0.8em}
{\@author\par}
\vspace{0.5em}
{\@date\par}
\end{center}
}
\makeatother
 
\titleformat{\section}
{\fontsize{12}{14}\bfseries}
{\thesection.}
{0.5em}
{\MakeUppercase}
\titleformat{\subsection}
{\fontsize{12}{14}\bfseries}
{\thesubsection}
{0.5em}
{}
\titleformat{\subsubsection}
{\fontsize{12}{14}\bfseries}
{\thesubsubsection}
{0.5em}
{}
 
\hypersetup{
    colorlinks=true,
    linkcolor=blue,
    citecolor=blue,
    urlcolor=blue
}
\title{Modeling the Dynamic Relationship Between Brent Crude Oil Prices and the Nepal Stock Exchange: An Integrated Econometric and Explainable Machine Learning Approach}
\author{
Anamol Khadka$^{1}$,
Milan Arjel$^{2}$,
Ayush Lataula$^{3}$,
Aayam Dhakal$^{4}$,
Prajun Trital$^{5}$,
Mingmar Sherpa$^{6}$,
Biman Rimal$^{7}$\\[8pt]
$^{1}$Department of Electronics and Computer Engineering,\\
Pulchowk Campus, Institute of Engineering, Tribhuvan University, Lalitpur, Nepal\\[4pt]
$^{2}$Department of Electronics and Computer Engineering, Purwanchal Campus, Institute of Engineering, Tribhuvan University, Dharan, Nepal \\[4pt]
$^{3}$Questrom School of Business,\\
Boston University, Massachusetts, United States of America\\[4pt]
$^{4}$Department of Chemical and Biomolecular Engineering,\\
Cornell University, New York, United States of America \\
$^{5}$Department of Computer Science,\\
University of Alabama in Huntsville, Alabama, United States of America\\
$^{6}$Department of Biomedical and Biological Sciences, Cornell University, New York, United States of America\\[4pt]
$^{7}$Department of Mechanical and Aerospace Engineering,\\
Institute of Engineering, Tribhuvan University, Lalitpur, Nepal\\[4pt]
}
\date{}
\begin{document}
\maketitle
 
\section{Corresponding Author : }
Mingmar Sherpa, Aayam Dhakal
\begin{abstract}
\raggedright
\noindent
This study examines the dynamic relationship between the global oil prices and Nepal Stock Exchange (NEPSE) using an integrated approach which combines traditional econometric techniques with machine learning and explainable AI techniques. For this, Daily data of International Oil prices and NEPSE index is analyzed from approximately thirteen years (June 2013 -- June 2026) using Granger causality, EGARCH(1,1), and DCC-GARCH models to examine different properties like predictive relationships, asymmetric volatility behaviour, and time-varying correlations. To further supplement the econometric analysis, Machine Learning Models like Random Forest, LightGBM, and XGBoost algorithms were used to capture nonlinear relationships, along with explainable artificial intelligence techniques like SHAP values, Partial Dependence Plots, and Individual Conditional Expectation plots to further interpret the results of the model. The  results from the econometric analysis showed a statistically significant unidirectional Granger causality from Brent crude oil to NEPSE with a four-day lag, high volatility persistence in both markets, and weak yet highly time-varying conditional correlations. Among the machine learning models, XGBoost achieves the best  performance, and explainability analysis reveals that NEPSE's own momentum and short-term volatility mainly influence its own behaviour and oil-related information serves as a minor, method-dependent contributor. The findings demonstrate that econometric and explainable machine learning approaches provide  insights into the oil--equity market relationship in a way that each approach complements the result of one another.
 
\medskip
\noindent\textbf{Keywords:} EGARCH; DCC-GARCH; XGBoost; SHAP; Explainable AI
\end{abstract}

\section{Introduction}

Crude oil is one of the most powerful commodities driving the global economy which shapes everything from production costs and inflation to exchange rates. For countries that rely heavily on oil imports like Nepal, these global price movements can intensively affect financial markets. This is exactly why investors and policymakers need a clear understanding of how oil prices interact with stock markets to make sound investment decisions and mitigate financial risk, keeping the economy stable.

Previous studies have examined the relationship between crude oil prices and
stock markets in a quantitative manner through econometric modeling. Sarwar et al. \cite{sarwar2018} investigated volatility
spillovers among major Asian oil-importing countries using multivariate GARCH models and
found that the magnitude and direction of volatility transmission differ across countries which
suggests that the oil--stock relationship is country-specific. Mushtaq et al. \cite{mushtaq2023} studied
South Asian stock markets  and reported significant long-run volatility spillovers from crude oil. In that study,
Nepal portrayed no significant short-run spillover but clear long-run effects. Their findings also
showed that the correlation between crude oil and stock markets changes over time,
highlighting the importance of dynamic modeling. More recently, Khalid et al. \cite{khalid2025} used
an EGARCH framework to assess the oil--stock relationship in ASEAN markets after the
COVID-19 pandemic. They found that crude oil continues to influence stock market volatility,
although the magnitude of spillovers has weakened compared with the pandemic period, and
that positive and negative oil price shocks affect stock market volatility differently.

Similarly, Xu et al. \cite{xu2024} expanded the volatility spillover literature by incorporating macroeconomic shocks into the analysis of interrelated factors across crude oil, equity, bond, and gold markets. Their findings demonstrate that volatility transmission is not only driven by oil price movements, but it also evolves according to broader macroeconomic conditions which suggests us that spillover relationships are dynamic rather than constant over time.

Further evidence supporting this dynamic perspective was provided by Ben Salem et al. \cite{bensalem2024}. They studied volatility spillovers between crude oil prices and major exchange rates using dynamic connectedness techniques. Their study focused on currency markets, but it still demonstrated that financial spillovers evolve continuously under changing economic conditions, reinforcing the importance of employing time-varying econometric models when analysing cross-market relationships.

While these studies have explored the volatility spillovers, asymmetric behaviours and shifting
correlations, they primarily used traditional econometric models that assume a pre-defined
functional behaviour about how stock returns behave relative to oil prices. Even though they
show that a relationship exists between them, the underlying functional forms and behaviour of
their relationship remain largely unexplored. In particular, Nepal has received very little focused
attention when it comes to understanding how oil price shocks play out in its stock exchange. 

To address this research gap, our study aims to integrate econometric models with machine
learning and Explainable AI approaches. Different econometric models are used to study
volatility spillovers and causal relationships between oil prices and the stock market. After that, three
ensemble learning algorithms (Random Forest, LightGBM, and XGBoost) are compared to model
the relationship between oil prices and stock market returns without assuming a predefined functional
form. The best-performing model is then explained using multiple explainability techniques:
feature importance rankings, SHAP (SHapley Additive exPlanations) values, Partial Dependence
Plots (PDP), and Individual Conditional Expectation (ICE) plots. SHAP values quantify each
feature's contribution to individual predictions, PDP reveals the average effect of key predictors,
and ICE examines how this relationship varies across observations, revealing non-linear patterns
and heterogeneous responses that conventional econometric models may not capture. Methodological studies like this will be very helpful in understanding the market dynamics of markets like NEPSE which have received very little attention in the international financial studies. Such integrated studies that utilize multiple approaches will help better understand the overall market trend in a much deeper manner.

\section{Material and Methods}

\subsection{Data and Research Framework}

This study has utilized a quantitative time series model to examine the dynamic relationship between global Brent crude oil prices and the Nepal Stock Exchange (NEPSE). An integrated analytical framework incorporating an explainable machine learning model was employed to study the relationship in a manner where each finding complements the conclusions of the other. The econometric component evaluated stationarity, predictive causality, volatility persistence, and time-varying conditional correlations. On the other hand, machine learning models were used to capture nonlinear relationships that may not adequately be represented by conventional parametric models. Finally, explainable artificial intelligence (XAI) methods were applied to interpret the relationships discovered and identify the contributions of different features to overall model predictions.

Daily time-series data which consists of Brent crude oil prices and the NEPSE Index were used for the analysis by the model. The data for the crude oil prices were obtained from the Federal Reserve Economic Data (FRED) database \cite{eia2026} and the historical NEPSE index data was compiled from the Mendeley dataset  \cite{khana2026} and extended using NepseAlpha post consistency validation during overlapping periods of the two datasets. The final dataset starts from 25 June 2013 and ends in 22 June 2026. It nearly provided  thirteen years of matched daily observations having 2,335 trading days.

\begin{table}[H]
\centering
\caption{Data Sources}
\label{tab:data_sources}
\begin{tabular}{lll}
\toprule
\textbf{Variable} & \textbf{Source} & \textbf{Citation} \\
\midrule
Brent Crude Oil (USD/barrel) & FRED \texttt{DCOILBRENTEU} & \cite{eia2026} \\
NEPSE Index (2004--2024) & Mendeley Data & \cite{khana2026} \\
NEPSE Index (2024--2026) & NepseAlpha & \cite{nepsealpha2026}\\
\bottomrule
\end{tabular}
\end{table}

Before the data analysis, repeated observations, invalid values, and missing records were removed. Furthermore, the date formats were standardized. As international oil markets and the NEPSE operate under different trading calendars, the datasets were synchronized using an inner join. It was done in such a way that only the common trading dates were kept and the artificial observations, that could possibly arise through interpolation or forward filling, were avoided.

Daily logarithmic returns were then calculated as:

\begin{equation}
R_t = \ln\!\left(\frac{P_t}{P_{t-1}}\right) \times 100
\label{eq:logreturns}
\end{equation}

\noindent where $P_t$ represents the closing price at time $t$. Logarithmic returns were used in this research because logarithmic returns generally satisfy the stationarity assumptions that are essential for financial econometric modelling.

\begin{table}[H]
\centering
\caption{Descriptive Statistics --- Price Levels}
\label{tab:desc_levels}
\begin{tabular}{lrr}
\toprule
\textbf{Statistic} & \textbf{Brent Crude Oil (USD)} & \textbf{NEPSE Index} \\
\midrule
Mean   & 72.57    & 1741.38 \\
Std. Dev. & 21.52 & 681.14 \\
Min    & 22.79\footnote{This represents the minimum Brent crude oil price within the filtered dataset of matched NEPSE trading days, rather than the absolute global market minimum.}    & 493.96  \\
Max    & 138.21   & 3199.03 \\
\bottomrule
\end{tabular}
\end{table}

\begin{table}[H]
\centering
\caption{Descriptive Statistics --- Logarithmic Returns (\%)}
\label{tab:desc_returns}
\begin{tabular}{lrr}
\toprule
\textbf{Statistic} & \textbf{Brent Return} & \textbf{NEPSE Return} \\
\midrule
Mean   & $-0.0114$ & 0.0732 \\
Std. Dev. & 2.9389 & 1.5074 \\
Min    & $-36.5993$\footnote{These extreme single-day returns correspond to historical volatility shocks during the April 2020 pandemic crash.} & $-8.7587$ \\
Max    & 39.9351   & 8.0504  \\
\bottomrule
\end{tabular}
\end{table}

For the machine learning analysis, additional features were engineered from the original crude oil and NEPSE time series. These features are lagged returns (1--5 trading days), rolling means and rolling standard deviations over 5-, 10-, and 22-day windows, squared and absolute returns, calendar variables (day of week, month, and year), and a 22-day rolling correlation between Brent crude oil and NEPSE returns. To prevent look-ahead bias, all rolling-window features and derivative metrics were constructed using a one-day lag before calculation. This ensures that predictions for day $t$ rely only on information available up to day $t-1$. Observations containing missing values due to lagging and rolling-window calculations were eliminated prior to model training, yielding 2,314 observations with 28 candidate features. After excluding four redundant rolling mean features (Brent\_Mean\_5d/10d/22d and NEPSE\_Mean\_22d) to reduce multicollinearity (see Section~3.5), the final predictor set comprised 24 features.

\subsection{Econometric Analysis}

The econometric analysis was performed one after another to examine the statistical properties, predictive relationships, and volatility dynamics between Brent crude oil prices and the Nepal Stock Exchange. The analytical framework consisted of stationarity testing, conditional heteroskedasticity diagnostics, Granger causality analysis, asymmetric volatility modelling using EGARCH, and estimation of dynamic conditional correlations through the DCC-GARCH model.

\subsubsection{Stationarity Tests}

The stationarity of Brent crude oil prices and the NEPSE index was tested using the Augmented Dickey--Fuller (ADF) and Phillips--Perron (PP) unit root tests. The ADF test studies serial correlation using lagged differences, whereas the PP test applies non-parametric corrections for serial correlation and heteroskedasticity.

The ADF regression is:
\begin{equation}
\Delta y_t = \alpha + \beta t + \gamma y_{t-1} + \sum_{i=1}^{p} \delta_i \Delta y_{t-i} + \varepsilon_t
\label{eq:adf}
\end{equation}

\noindent with hypotheses:
\begin{align}
H_0 &: \gamma = 0 \notag \\
H_1 &: \gamma < 0 \notag
\end{align}

The PP test estimates:
\begin{equation}
\Delta y_t = \alpha + \beta t + \gamma y_{t-1} + \varepsilon_t
\label{eq:pp}
\end{equation}

\noindent while correcting the test statistics using Newey--West estimators.

\subsubsection{ARCH-LM Test}

The ARCH-LM test was used to determine whether the return series show conditional heteroskedasticity. Rejection of the null hypothesis indicates clustering of volatility and justifies the use of GARCH-family models in this study statistically.

\subsubsection{Granger Causality Analysis}

Granger causality analysis was done to examine whether historical Brent crude oil returns contain predictive information about future NEPSE returns and vice versa.

For NEPSE:
\begin{equation}
\text{NEPSE}_t = \alpha + \sum_{i=1}^{p} \beta_i \, \text{NEPSE}_{t-i} + \sum_{i=1}^{p} \gamma_i \, \text{Oil}_{t-i} + \varepsilon_t
\label{eq:granger_nepse}
\end{equation}

For Brent crude oil:
\begin{equation}
\text{Oil}_t = \alpha + \sum_{i=1}^{p} \beta_i \, \text{Oil}_{t-i} + \sum_{i=1}^{p} \gamma_i \, \text{NEPSE}_{t-i} + \varepsilon_t
\label{eq:granger_oil}
\end{equation}

The null hypothesis is:
\begin{equation}
H_0 : \gamma_1 = \gamma_2 = \cdots = \gamma_p = 0
\notag
\end{equation}

\subsubsection{EGARCH Model}

To study asymmetric volatility behaviour, an EGARCH(1,1) model was used \cite{nelson1991}:

\begin{equation}
\ln(\sigma_t^2) = \omega + \alpha \left|\frac{\varepsilon_{t-1}}{\sigma_{t-1}}\right| + \gamma \left(\frac{\varepsilon_{t-1}}{\sigma_{t-1}}\right) + \beta \ln(\sigma_{t-1}^2)
\label{eq:egarch}
\end{equation}

\noindent where $\gamma$ captures the leverage effect and $\beta$ measures volatility persistence.

\subsubsection{DCC-GARCH Model}

Dynamic conditional correlations between Brent crude oil returns and NEPSE returns were estimated using the DCC-GARCH framework \cite{engle2002}.

The conditional covariance matrix is:
\begin{equation}
H_t = D_t R_t D_t
\label{eq:dcc_cov}
\end{equation}

\noindent where the dynamic covariance process is:
\begin{equation}
Q_t = (1 - a - b)\bar{Q} + a(z_{t-1} z_{t-1}') + b Q_{t-1}
\label{eq:dcc_q}
\end{equation}

\noindent and the correlation matrix is:
\begin{equation}
R_t = Q_t^{*-1} Q_t Q_t^{*-1}
\label{eq:dcc_r}
\end{equation}

\subsection{Nonlinear Relationship Modelling}

To further extend the conclusions of the econometric analysis, three different learning algorithms (Random Forest \cite{breiman2001}, LightGBM \cite{ke2017}, and XGBoost \cite{chen2016}) were used to model the nonlinear relationships between Brent crude oil prices and NEPSE returns. Model training and development was based primarily on features engineered from the Brent and NEPSE time series. It ensured that the consistency with the econometric framework and avoided the introduction of external macroeconomic variables. To reduce multicollinearity and improve generalization ability of the model, redundant rolling mean features (Brent\_Mean\_5d/10d/22d and NEPSE\_Mean\_22d) were excluded from the predictor set, as Brent rolling means add no signal beyond individual lagged returns in an efficiently traded global market, and NEPSE\_Mean\_22d is highly correlated with the retained short- and medium-term momentum features.

During Model development, an 80:20 chronological train--test split was used to ensure that all observations in the test set occurred after the training period. Hyperparameter selection emphasised regularization to limit overfitting and improve out-of-sample generalization Parameters such as tree depth, learning rate, subsampling ratios, and L1/L2 penalties were constrained. Likewise, Five-fold expanding-window TimeSeriesSplit cross-validation was used on the training data to verify stability. After selecting the hyperparameters, each model was retrained on the full training set. The model was also evaluated using Root Mean Squared Error (RMSE), Mean Absolute Error (MAE), the coefficient of determination ($R^2$), and Directional Accuracy (DA). The best-performing model was then selected for explainability analysis.

The evaluation metrics are:

\begin{equation}
\text{RMSE} = \sqrt{\frac{1}{n} \sum_{i=1}^{n} (y_i - \hat{y}_i)^2}
\label{eq:rmse}
\end{equation}

\begin{equation}
\text{MAE} = \frac{1}{n} \sum_{i=1}^{n} |y_i - \hat{y}_i|
\label{eq:mae}
\end{equation}

\begin{equation}
R^2 = 1 - \frac{\sum (y_i - \hat{y}_i)^2}{\sum (y_i - \bar{y})^2}
\label{eq:r2}
\end{equation}

\begin{equation}
\text{DA} = \frac{1}{n} \sum \mathbb{I}\!\left(\text{sign}(y_i) = \text{sign}(\hat{y}_i)\right) \times 100
\label{eq:da}
\end{equation}

\subsection{Explainable Artificial Intelligence}

Explainable artificial intelligence techniques were applied to the best-performing machine learning model to improve interpretability. Feature importance analysis was first conducted to identify the most influential predictors. SHapley Additive exPlanations (SHAP) \cite{lundberg2017} were then employed to quantify the contribution of individual features to each prediction. Partial Dependence Plots (PDP) were used to visualize average nonlinear effects, while Individual Conditional Expectation (ICE) plots examined heterogeneous responses across individual observations. Together, these techniques provide both global and local interpretations of the learned relationship between Brent crude oil prices and the Nepal Stock Exchange.

\section{Results}

\subsection{Preliminary Analysis}

The statistical properties of the Brent crude oil and NEPSE series were first examined before using the econometric and machine learning models. Unit root tests indicate that both Brent crude oil prices and the NEPSE index are non-stationary in levels but become stationary after first differencing through logarithmic returns. Both the Augmented Dickey--Fuller (ADF) and Phillips--Perron (PP) tests consistently classify the price series as integrated of order one, while the corresponding return series are stationary at the 1\% significance level. These findings are consistent with the behaviour usually observed in financial time series and confirm that return series are appropriate for subsequent econometric analysis.

\begin{table}[H]
\centering
\caption{Unit Root Test Results}
\label{tab:unitroot}
\begin{tabular}{lcccc}
\toprule
\textbf{Variable} & \textbf{ADF Statistic} & \textbf{ADF $p$-value} & \textbf{PP Statistic} & \textbf{PP $p$-value} \\
\midrule
Brent Price (Level)  & $-2.8608$ & 0.0501 & $-2.5702$ & 0.0993 \\
NEPSE Index (Level)  & $-1.5903$ & 0.4885 & $-1.4918$ & 0.5376 \\
Brent Return         & $-18.3429$ & 0.0000 & $-46.2597$ & 0.0000 \\
NEPSE Return         & $-14.6575$ & 0.0000 & $-45.9342$ & 0.0000 \\
\bottomrule
\multicolumn{5}{l}{\footnotesize Critical values: 1\% = $-3.4332$, 5\% = $-2.8628$, 10\% = $-2.5674$} \\
\end{tabular}
\end{table}

The ARCH-LM test further revealed statistically significant ARCH effects in both Brent crude oil and NEPSE return series ($p < 0.001$), indicating the presence of volatility clustering. Consequently, the assumption of constant variance is violated, providing empirical justification for the use of GARCH-family models to examine volatility dynamics.

\begin{table}[H]
\centering
\caption{ARCH-LM Test Results}
\label{tab:archlm}
\begin{tabular}{lccc}
\toprule
\textbf{Variable} & \textbf{LM Statistic} & \textbf{$p$-value} & \textbf{Verdict} \\
\midrule
Brent Return & 425.4073 & 0.0000 & ARCH effects present \\
NEPSE Return & 160.9386 & 0.0000 & ARCH effects present \\
\bottomrule
\end{tabular}
\end{table}

\begin{figure}[H]
\centering
\includegraphics[width=\textwidth]{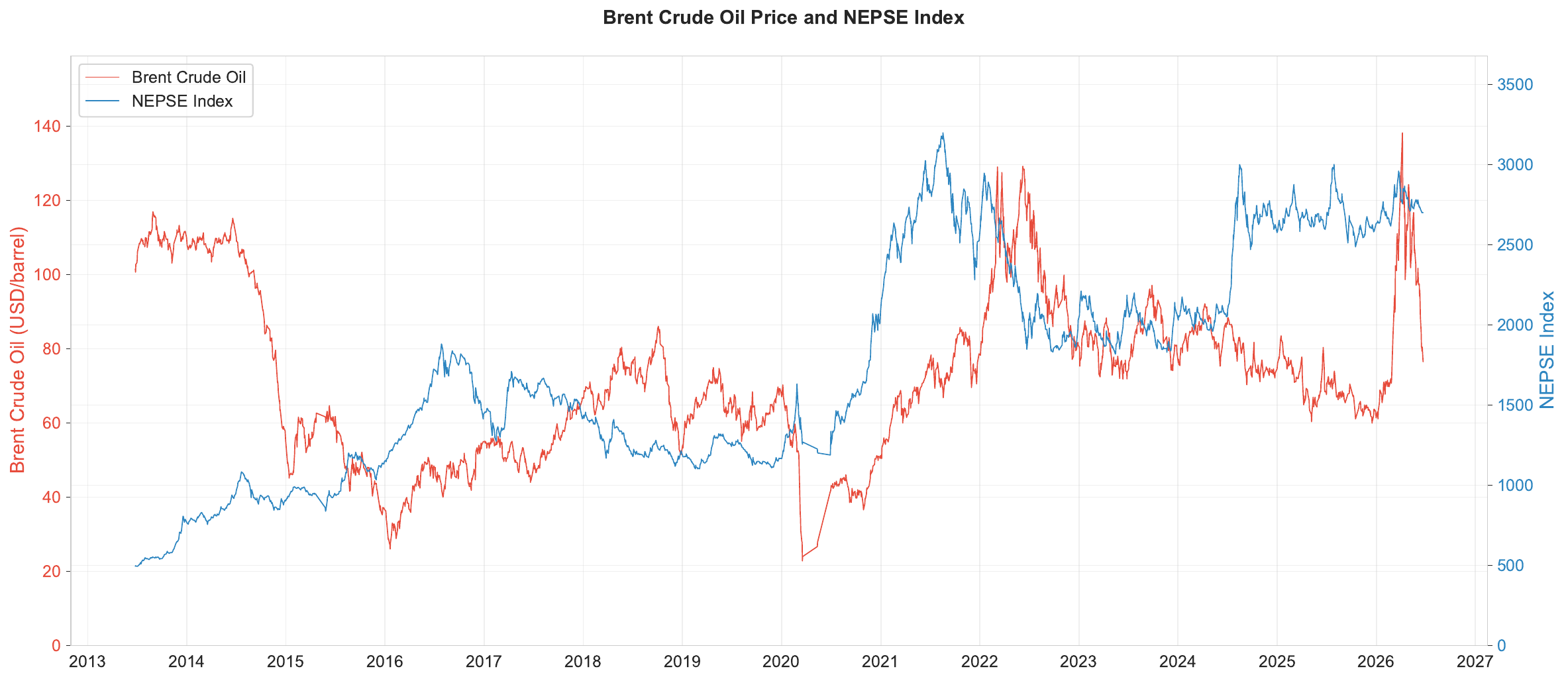}
\caption{Price level series of Brent crude oil and the NEPSE index (June 2013 -- June 2026).}
\label{fig:price_levels}
\end{figure}

\begin{figure}[H]
\centering
\includegraphics[width=\textwidth]{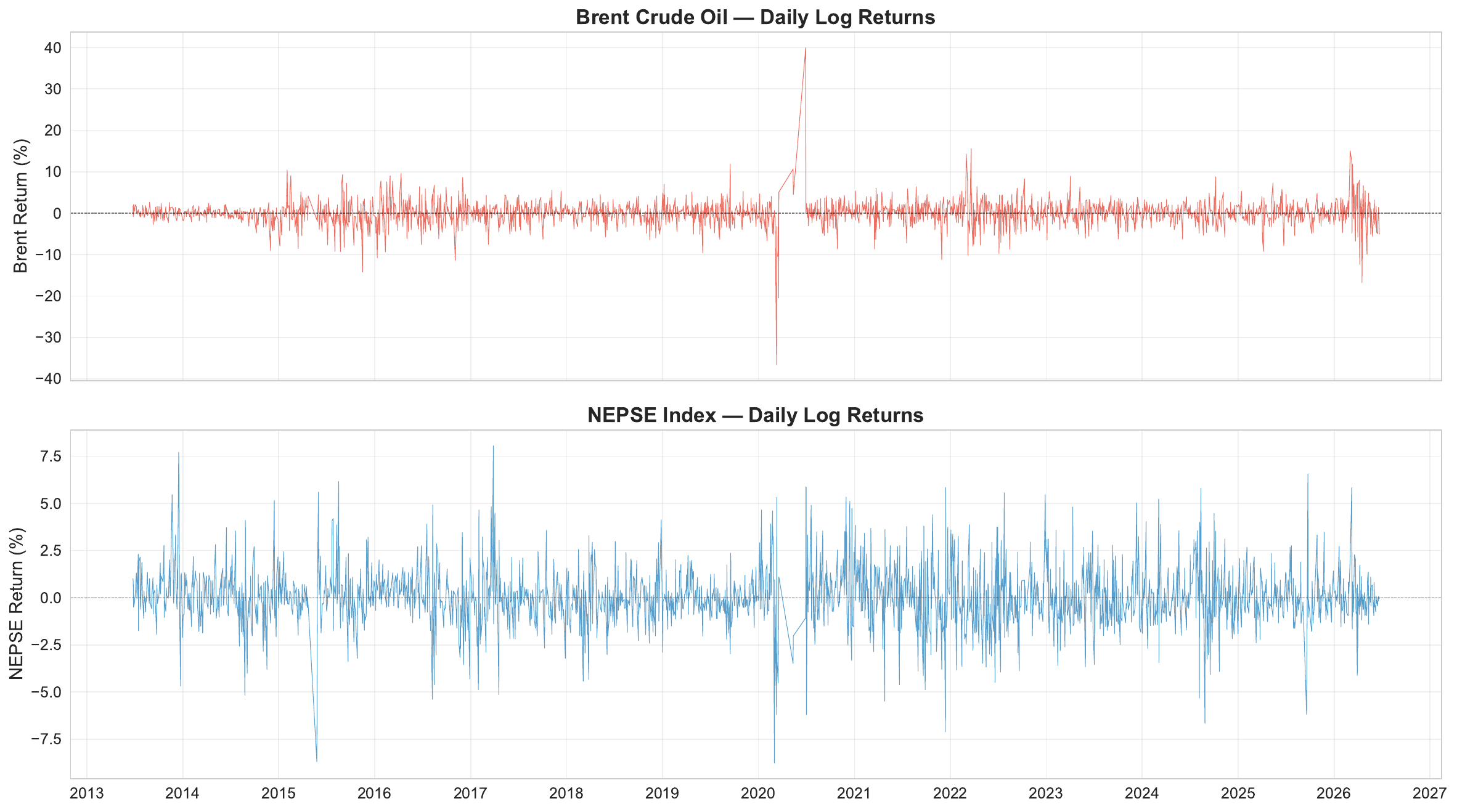}
\caption{Daily logarithmic return series for Brent crude oil and the NEPSE index.}
\label{fig:return_series}
\end{figure}

\begin{figure}[H]
\centering
\includegraphics[width=0.9\textwidth]{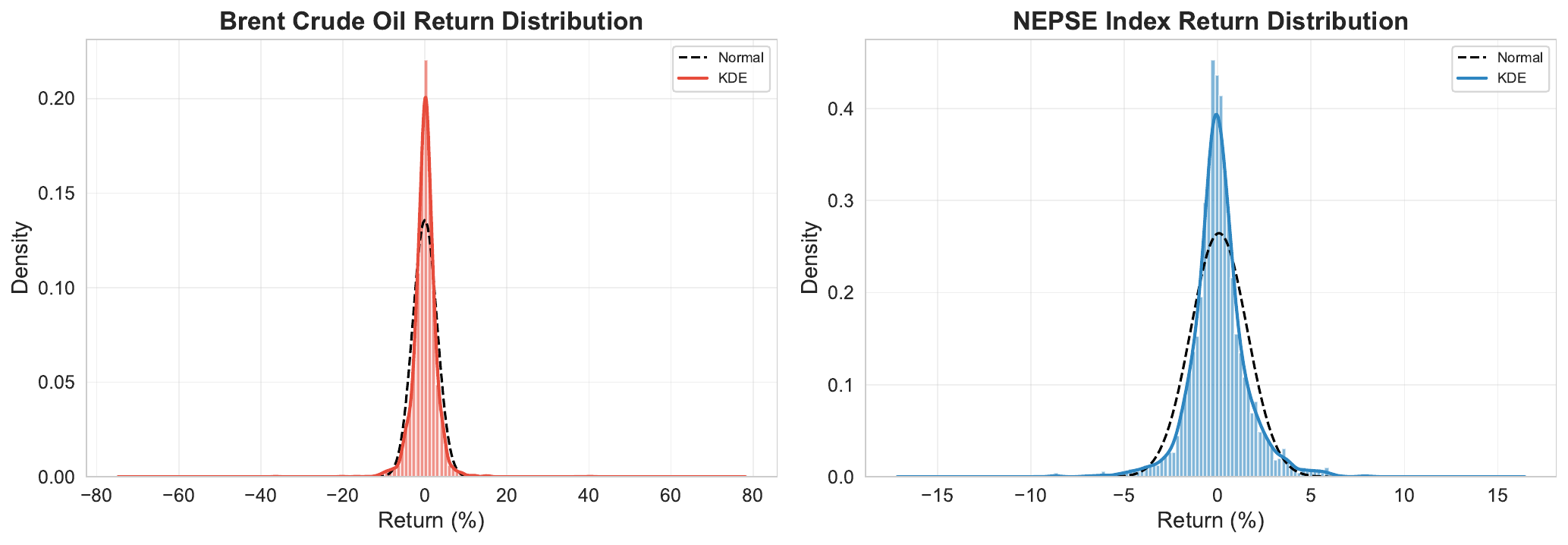}
\caption{Distribution of daily logarithmic returns for Brent crude oil and NEPSE.}
\label{fig:return_dist}
\end{figure}

\subsection{Predictive Relationship Between Brent Crude Oil and NEPSE}

Granger causality analysis was performed to determine whether historical movements in Brent crude oil prices contain predictive information for future movements in the Nepal Stock Exchange. The results indicate a statistically significant unidirectional relationship from Brent crude oil to NEPSE at a lag length of four trading days ($F = 2.5231$, $p = 0.0392$). On the other hand, no statistically significant evidence was found that NEPSE returns Granger-cause Brent crude oil prices.

\begin{table}[H]
\centering
\caption{Granger Causality Test Results}
\label{tab:granger}
\begin{tabular}{lcccc}
\toprule
\textbf{Direction} & \textbf{Best Lag} & \textbf{$F$-Statistic} & \textbf{$p$-value} & \textbf{Significant?} \\
\midrule
Oil $\rightarrow$ NEPSE & 4 & 2.5231 & 0.0392 & Yes (at 5\%) \\
NEPSE $\rightarrow$ Oil & 4 & 1.8350 & 0.1194 & No \\
\bottomrule
\end{tabular}
\end{table}

\begin{table}[H]
\centering
\caption{Granger Causality: Brent Oil $\rightarrow$ NEPSE (Lag 1--10)}
\label{tab:granger_detail}
\begin{tabular}{cccc}
\toprule
\textbf{Lag} & \textbf{$F$-Statistic} & \textbf{$p$-value} & \textbf{Significance} \\
\midrule
1  & 0.0538 & 0.8167 &  \\
2  & 1.9827 & 0.1379 &  \\
3  & 1.3805 & 0.2469 &  \\
4  & 2.5231 & 0.0392 & ** \\
5  & 2.0256 & 0.0721 & * \\
6  & 1.6929 & 0.1187 &  \\
7  & 1.7414 & 0.0951 & * \\
8  & 1.6488 & 0.1061 &  \\
9  & 1.5981 & 0.1101 &  \\
10 & 1.4437 & 0.1548 &  \\
\bottomrule
\multicolumn{4}{l}{\footnotesize *** $p<0.01$, ** $p<0.05$, * $p<0.10$} \\
\end{tabular}
\end{table}

These findings suggest that global crude oil prices contain incremental information about subsequent movements in the Nepalese equity market, whereas Nepal's relatively small capital market has no measurable influence on international oil prices. The calculated four-day transmission lag further indicates that the effect of external energy market information is not immediately reflected in NEPSE but propagates gradually through the domestic market.

\subsection{Volatility Dynamics}

The EGARCH(1,1) model was estimated to investigate whether Brent crude oil and NEPSE exhibit asymmetric volatility behaviour. For Brent crude oil, the persistence parameter was estimated at $\beta = 0.9856$, indicating that volatility shocks decay slowly over time. The asymmetry parameter ($\gamma = -0.0495$) was negative but only marginally significant ($p = 0.0603$), providing weak evidence that negative oil price shocks generate larger volatility responses than positive shocks of comparable magnitude.

For the Nepal Stock Exchange, volatility persistence also remained high ($\beta = 0.9169$). This suggeststha t periods of elevated market risk continue beyond the initial shock. However, the asymmetry parameter was statistically insignificant ($\gamma = -0.0097$, $p = 0.6384$), which implies that positive and negative return shocks show similar effects on NEPSE volatility. Overall, the results indicate that while both markets exhibit persistent volatility, asymmetric responses are more dominant in the global oil market than in Nepal's equity market.

\begin{table}[H]
\centering
\caption{EGARCH(1,1) Parameter Estimates}
\label{tab:egarch}
\begin{tabular}{lcccc}
\toprule
\textbf{Parameter} & \multicolumn{2}{c}{\textbf{Brent Crude Oil}} & \multicolumn{2}{c}{\textbf{NEPSE Index}} \\
\cmidrule(lr){2-3} \cmidrule(lr){4-5}
 & Coefficient & $p$-value & Coefficient & $p$-value \\
\midrule
$\omega$ (constant)       & 0.0420  & 0.0000 & 0.0789  & 0.0026 \\
$\alpha$ (ARCH term)      & 0.1654  & 0.0000 & 0.2275  & 0.0000 \\
$\gamma$ (asymmetry)      & $-0.0495$ & 0.0603 & $-0.0097$ & 0.6384 \\
$\beta$ (persistence)     & 0.9856  & 0.0000 & 0.9169  & 0.0000 \\
\midrule
Log-Likelihood            & \multicolumn{2}{c}{$-5508.74$} & \multicolumn{2}{c}{$-4117.21$} \\
AIC                       & \multicolumn{2}{c}{11029.47} & \multicolumn{2}{c}{8246.43} \\
BIC                       & \multicolumn{2}{c}{11064.01} & \multicolumn{2}{c}{8280.96} \\
\bottomrule
\end{tabular}
\end{table}

\begin{figure}[H]
\centering
\includegraphics[width=\textwidth]{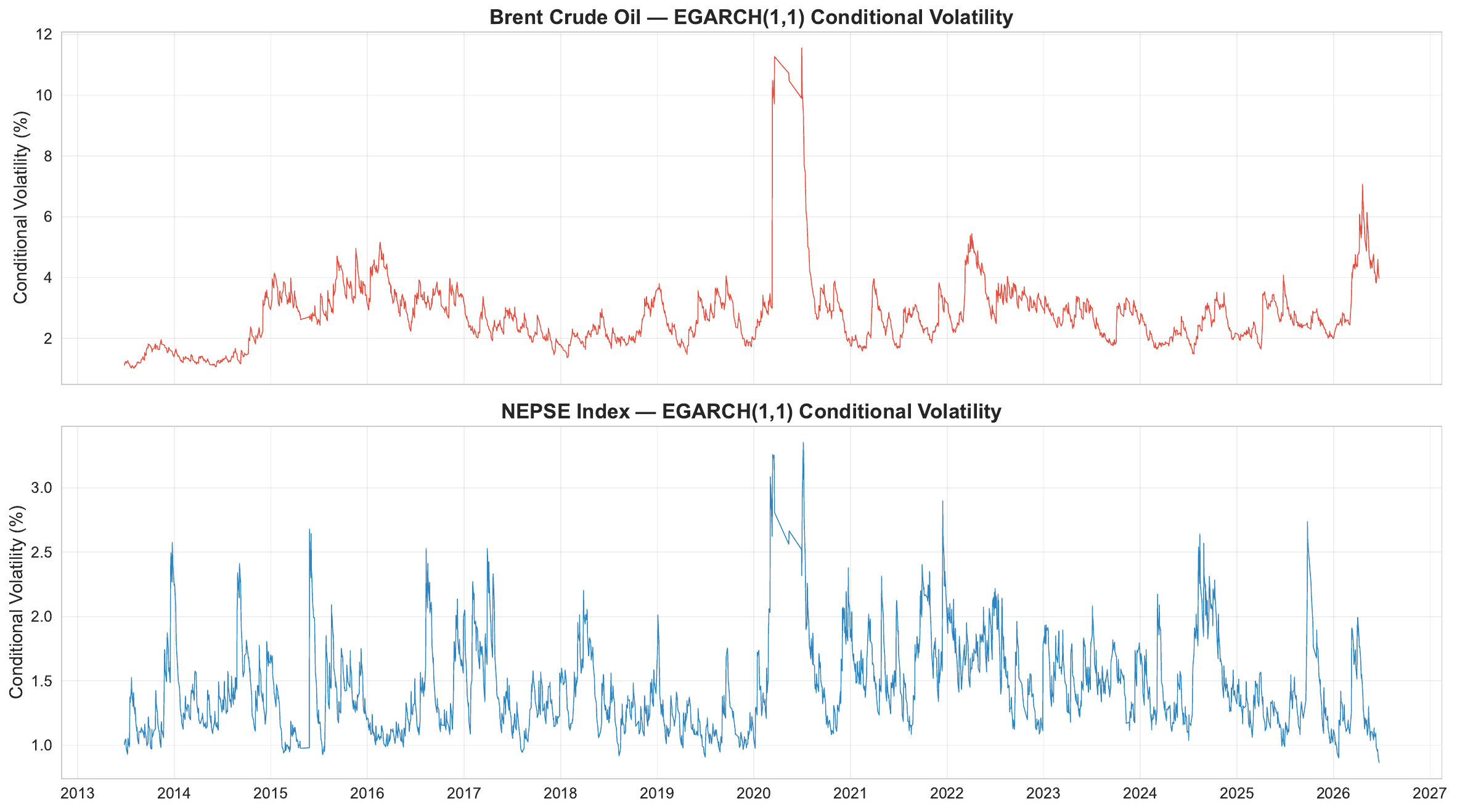}
\caption{EGARCH(1,1) conditional volatility for Brent crude oil and NEPSE.}
\label{fig:egarch}
\end{figure}

\subsection{Dynamic Conditional Correlation}

The DCC-GARCH model was employed to estimate the time-varying conditional correlation between Brent crude oil returns and NEPSE returns. The estimated average dynamic conditional correlation was 0.0168, which indicates an almost negligible linear association between the two markets on average. However, the correlation fluctuated considerably over time, ranging from $-0.3036$ to $0.3779$.

\begin{table}[H]
\centering
\caption{DCC-GARCH Estimation Results}
\label{tab:dcc}
\begin{tabular}{lr}
\toprule
\textbf{Parameter / Metric} & \textbf{Value} \\
\midrule
$a$ (news impact)    & 0.0390 \\
$b$ (persistence)    & 0.5000 \\
$a + b$              & 0.5390 \\
\midrule
Mean DCC             & 0.0168 \\
Std. Dev. DCC        & 0.0466 \\
Min DCC              & $-0.3036$ \\
Max DCC              & 0.3779 \\
\bottomrule
\end{tabular}
\end{table}

These findings demonstrate that the relationship between Brent crude oil and NEPSE is not constant. Instead, periods of positive co-movement alternate with periods of negative correlation. It suggests that external oil market shocks influence the Nepalese equity market only under particular market conditions. Also, static correlation measures may underestimate the complexity of the oil--NEPSE relationship, and the DCC framework captures its evolving nature more effectively.

\begin{figure}[H]
\centering
\includegraphics[width=\textwidth]{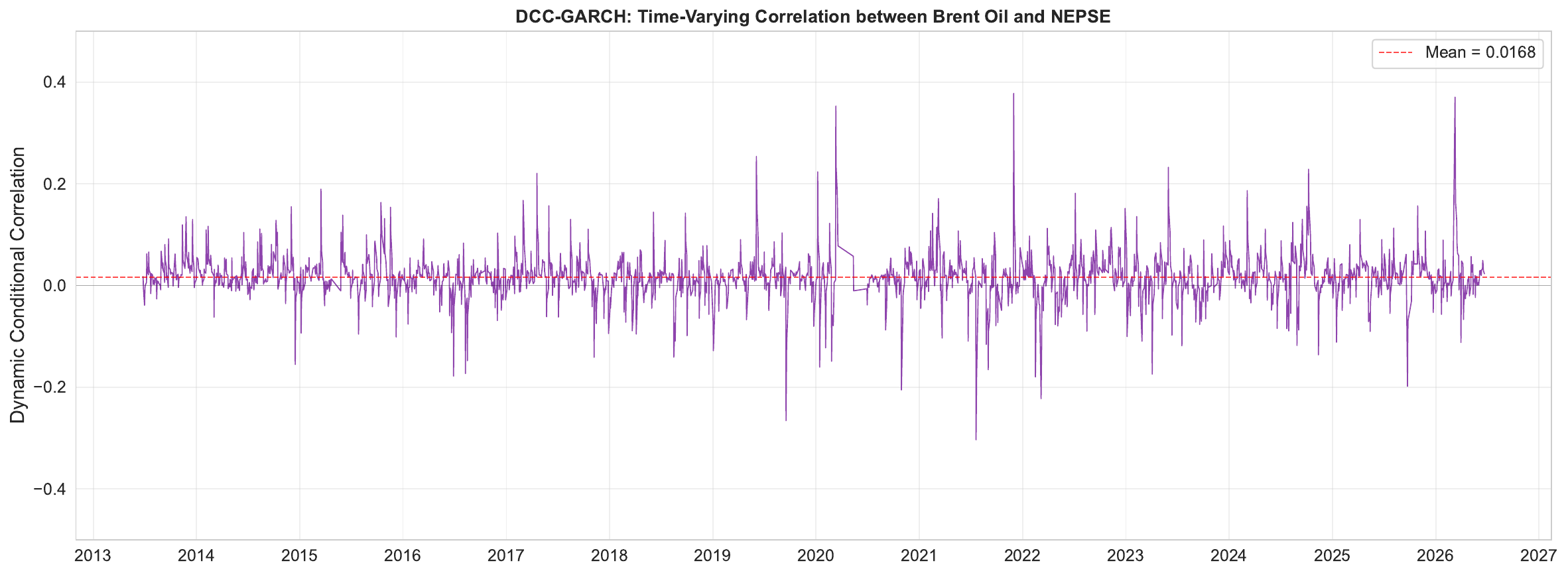}
\caption{DCC-GARCH: Time-varying conditional correlation between Brent crude oil and NEPSE returns.}
\label{fig:dcc}
\end{figure}

\begin{figure}[H]
\centering
\includegraphics[width=\textwidth]{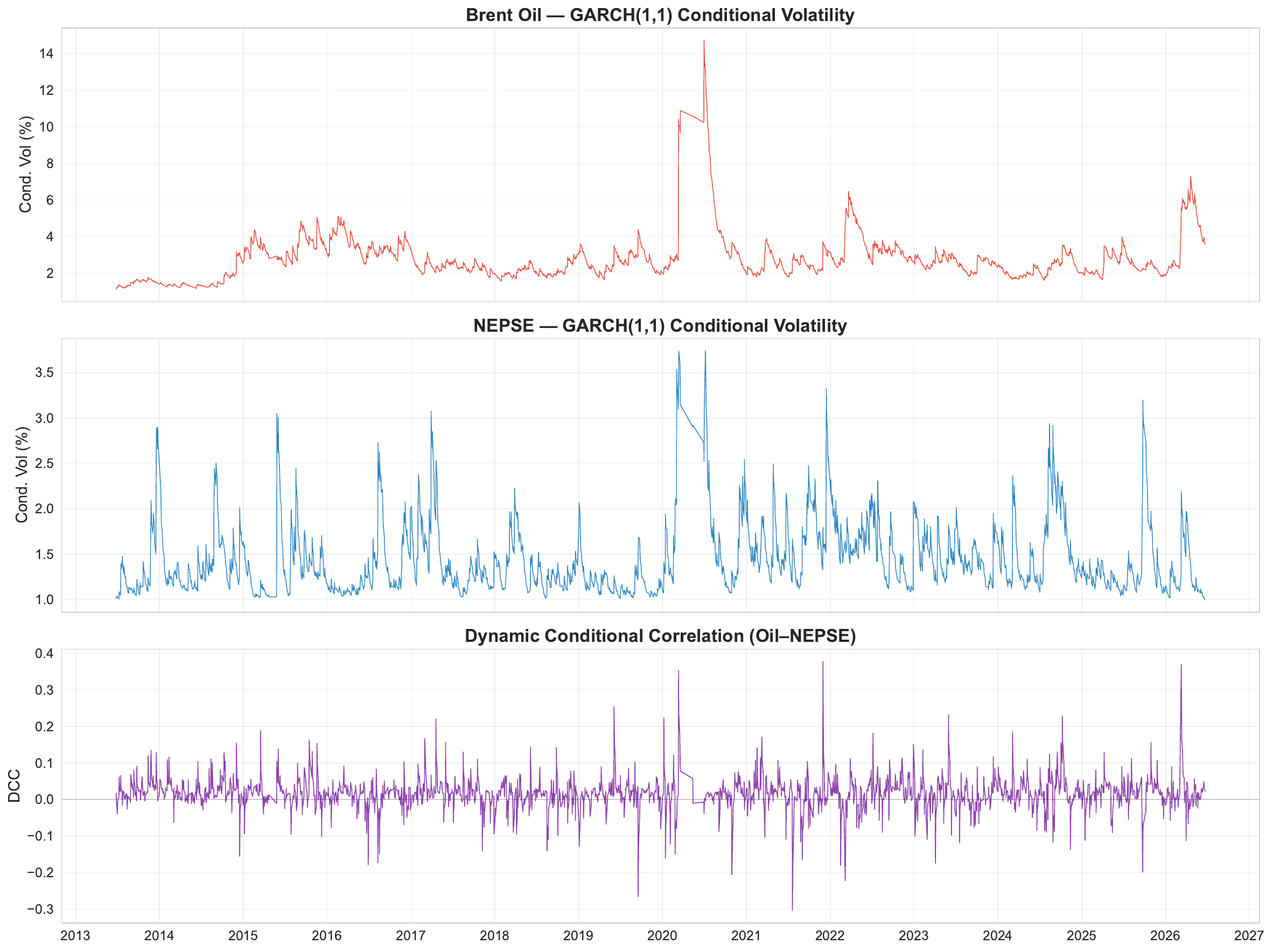}
\caption{DCC-GARCH full panel: conditional volatilities and dynamic correlation.}
\label{fig:dcc_panel}
\end{figure}

\begin{figure}[H]
\centering
\includegraphics[width=\textwidth]{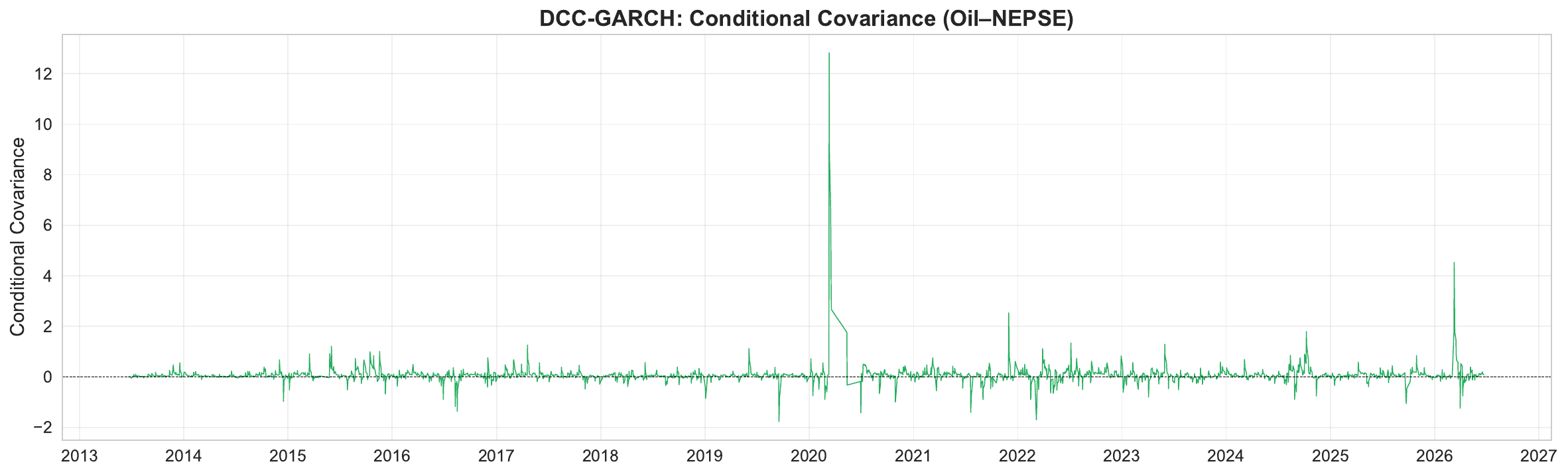}
\caption{DCC-GARCH: Conditional covariance between Brent crude oil and NEPSE returns.}
\label{fig:cond_cov}
\end{figure}

\subsection{Nonlinear Relationship Modelling}

The predictive performance of three ensemble learning algorithms --- Random Forest, LightGBM, and XGBoost --- was evaluated using an 80:20 temporal train--test split with five-fold TimeSeriesSplit cross-validation. Model performance was assessed using RMSE, MAE, $R^2$, and Directional Accuracy.

\begin{table}[H]
\centering
\caption{Machine Learning Model Comparison --- Out-of-Sample Test Set}
\label{tab:ml_comparison}
\begin{tabular}{lcccc}
\toprule
\textbf{Model} & \textbf{RMSE} & \textbf{MAE} & \textbf{$R^2$} & \textbf{DA (\%)} \\
\midrule
Random Forest (Bagging)       & 1.2176 & 0.8373 & 0.2838 & 66.3 \\
LightGBM (Gradient Boosting)  & 1.2901 & 0.8724 & 0.1961 & 63.9 \\
XGBoost (Gradient Boosting)   & \textbf{1.2119} & \textbf{0.8204} & \textbf{0.2905} & \textbf{67.0} \\
\bottomrule
\end{tabular}
\end{table}

\begin{table}[H]
\centering
\caption{Regularized Hyperparameters}
\label{tab:hyperparams}
\begin{tabular}{lp{10cm}}
\toprule
\textbf{Model} & \textbf{Parameters} \\
\midrule
Random Forest & \texttt{n\_estimators=500, max\_depth=4, min\_samples\_leaf=30} \\
LightGBM      & \texttt{n\_estimators=500, max\_depth=2, learning\_rate=0.003, subsample=0.7} \\
XGBoost       & \texttt{n\_estimators=300, max\_depth=3, learning\_rate=0.008, subsample=0.7, reg\_alpha=1.0} \\
\bottomrule
\end{tabular}
\end{table}

Among the evaluated models, XGBoost had the best predictive performance with an RMSE of 1.2119, an MAE of 0.8204, an $R^2$ of 0.2905, and a Directional Accuracy of 67.0\%. Just behind, Random Forest had a $R^2$ of 0.2838 and a DA of 66.3\%, while LightGBM, which is limited by its shallower tree depth, explained approximately 2\% of the variance with a DA of 63.9\%. All three testedmodels  outperformed the na\"ive zero-return benchmark (RMSE~=~1.4408), which confirms that meaningful predictive signal exists in the feature set. The Diebold--Mariano test indicates that XGBoost is statistically significantly better than LightGBM ($p < 0.001$). However, the same test reveals that  the difference between XGBoost and Random Forest is not statistically significant ($p = 0.77$). This suggests both gradient-boosted and bagging ensembles capture similar trends  in similar dynamics.

These results require careful and thorough interpretation. An $R^2$ of approximately 0.29 and a directional accuracy of 67\% for daily equity returns exceed the near-zero predictability typically reported for daily-return forecasting in developed markets \cite{campbell2008}. This performance does not necessarily help in economically meaningful trading opportunities after taking  transaction costs and different market frictions into consideration. The moderate performance is consistent with the characteristics of a small yet growing market like NEPSE, where non-synchronous trading \cite{lo1990} and lower liquidity create short-term serial correlation at the index level. This effect is reflected in the dominance of NEPSE\_Mean\_5d in the SHAP rankings (Table~\ref{tab:shap_importance}), where its mean absolute SHAP value is over three times that of the next-ranked feature. Rather than indicating that NEPSE returns are easily predictable, these results should be read as reflecting index-level microstructure effects combined with genuine short-horizon momentum. Five-fold TimeSeriesSplit cross-validation confirms the stability of these findings, with mean CV $R^2 = 0.28$ and mean DA of 66.5\% across expanding windows.
\begin{figure}[H]
\centering
\includegraphics[width=\textwidth]{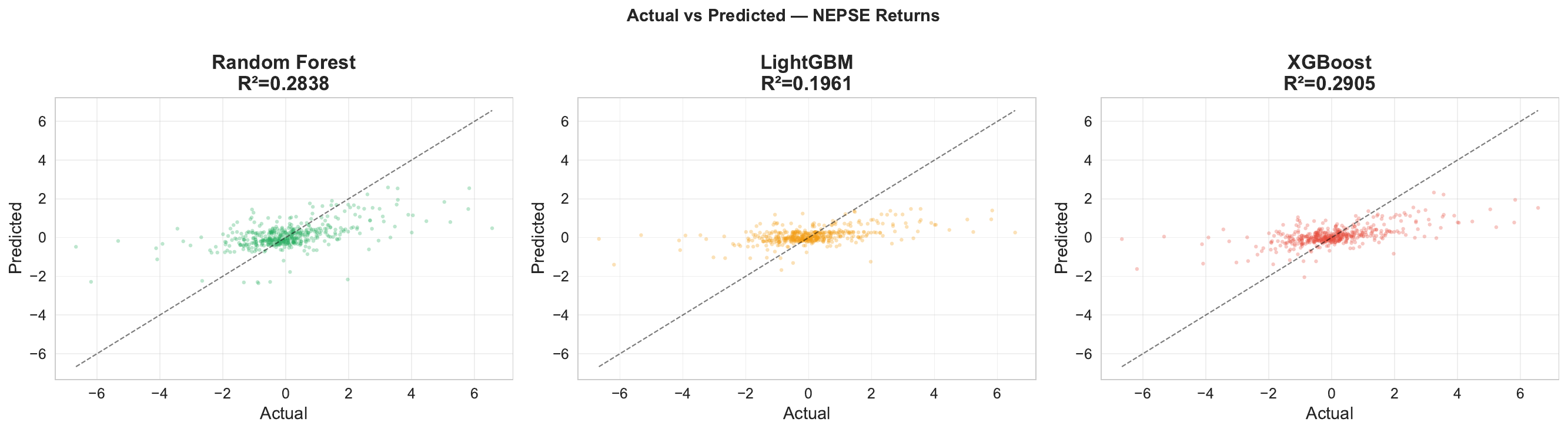}
\caption{Actual vs. predicted NEPSE returns for all three machine learning models.}
\label{fig:ml_scatter}
\end{figure}

\begin{figure}[H]
\centering
\includegraphics[width=\textwidth]{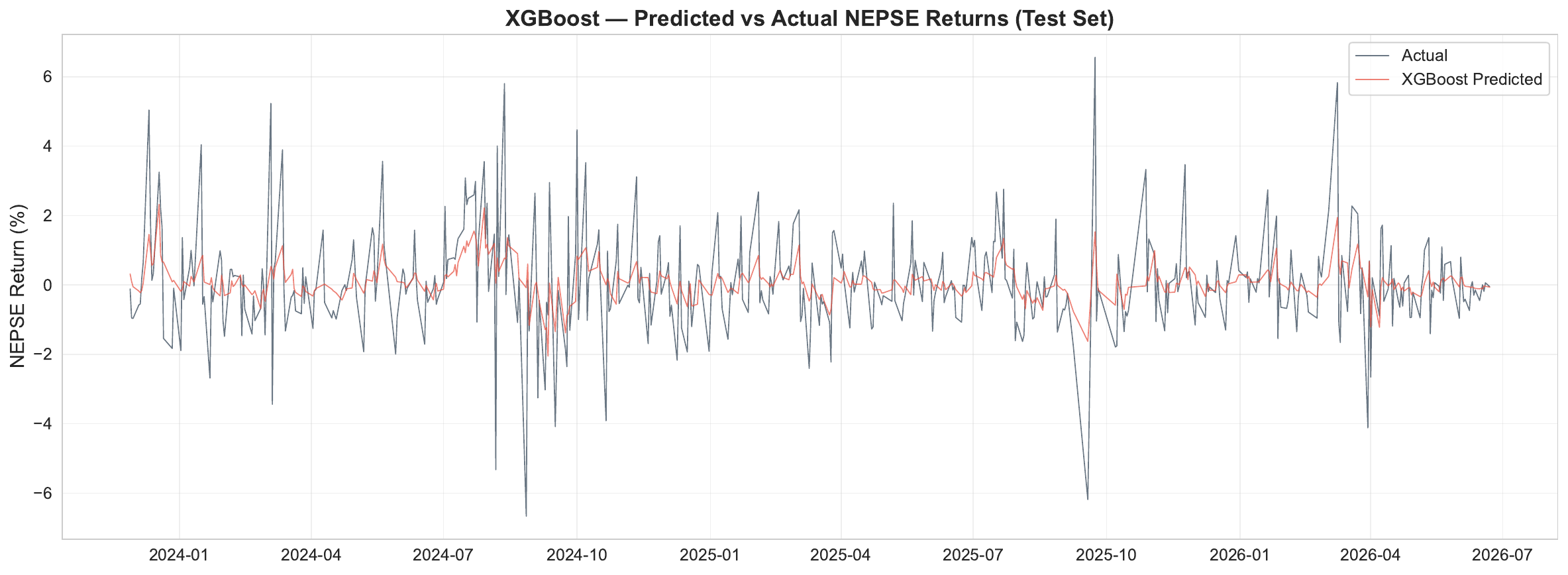}
\caption{XGBoost predicted vs. actual NEPSE returns over the test period.}
\label{fig:ml_ts}
\end{figure}

\begin{figure}[H]
\centering
\includegraphics[width=0.75\textwidth]{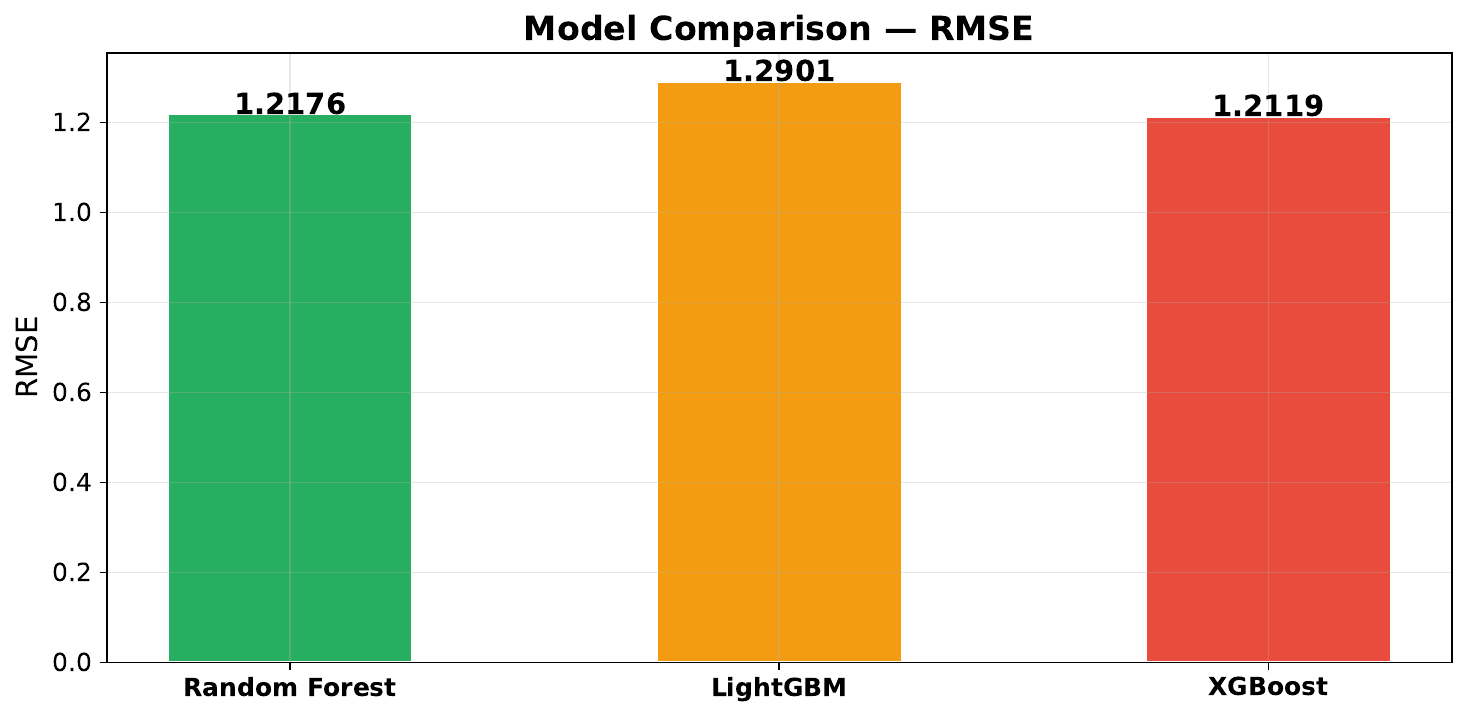}
\caption{RMSE comparison across machine learning models.}
\label{fig:ml_rmse}
\end{figure}

\subsection{Explainable Artificial Intelligence}

To improve the interpretation capability of the model, explainability analyses were conducted on the XGBoost model only as it better captured the non-linear characteristics compared to other models. Feature importance rankings indicate that variables derived from the NEPSE series dominate model predictions. In particular, NEPSE\_Mean\_5d and NEPSE\_Mean\_10d are the two most important predictors, followed by DayOfWeek and recent lagged NEPSE returns. The squared Brent return (Brent\_Return\_sq) appears within the top ten gain-based features, although its contribution is small relative to the NEPSE-derived predictors.

\begin{table}[H]
\centering
\caption{XGBoost Feature Importance --- Top 10 Features}
\label{tab:feature_importance}
\begin{tabular}{clcl}
\toprule
\textbf{Rank} & \textbf{Feature} & \textbf{Importance} & \textbf{Category} \\
\midrule
1  & NEPSE\_Mean\_5d     & 0.1323 & NEPSE momentum \\
2  & NEPSE\_Mean\_10d    & 0.1018 & NEPSE momentum \\
3  & DayOfWeek           & 0.0634 & Calendar \\
4  & NEPSE\_Return\_lag1 & 0.0535 & NEPSE autoregressive \\
5  & NEPSE\_Return\_lag2 & 0.0523 & NEPSE autoregressive \\
6  & NEPSE\_Vol\_10d     & 0.0451 & NEPSE volatility regime \\
7  & NEPSE\_Return\_lag4 & 0.0437 & NEPSE autoregressive \\
8  & NEPSE\_Vol\_5d      & 0.0435 & NEPSE short-term volatility \\
9  & NEPSE\_Return\_lag3 & 0.0434 & NEPSE autoregressive \\
10 & Brent\_Return\_sq   & 0.0423 & Oil signal \\
\bottomrule
\end{tabular}
\end{table}

It is worth noting that Brent\_Return\_sq appears in the gain-based top ten (Table~\ref{tab:feature_importance}) but does not appear in the SHAP top ten (Table~\ref{tab:shap_importance}), where all ten positions are occupied by NEPSE-derived features and DayOfWeek. This discrepancy indicates that the oil signal captured by the gain metric is small and method-dependent; once marginal contributions are measured via SHAP, the oil-related features fall below the reporting threshold. Oil's contribution to the model should therefore be characterised as minor rather than meaningful.

SHAP analysis confirms that NEPSE\_Mean\_5d is the most influential feature by mean absolute SHAP value (0.2983), reflecting strong short-term momentum effects in the Nepalese equity market. NEPSE\_Mean\_10d ranks second (0.0849), followed by several lagged NEPSE returns. Higher values of NEPSE\_Mean\_5d generally increase predicted returns, whereas elevated short-term volatility increases prediction uncertainty. Partial Dependence Plots show a largely positive relationship between recent market momentum and predicted returns, while volatility-related variables exhibit clear nonlinear behaviour. Individual Conditional Expectation plots additionally reveal heterogeneous responses across market conditions, which suggests that the influence of volatility depends on the prevailing market regime rather than remaining constant across observations.

\begin{table}[H]
\centering
\caption{Top 10 Features by Mean Absolute SHAP Value}
\label{tab:shap_importance}
\begin{tabular}{clcl}
\toprule
\textbf{Rank} & \textbf{Feature} & \textbf{Mean $|$SHAP$|$} & \textbf{Category} \\
\midrule
1  & NEPSE\_Mean\_5d      & 0.2983 & NEPSE momentum \\
2  & NEPSE\_Mean\_10d     & 0.0849 & NEPSE momentum \\
3  & NEPSE\_Return\_lag2  & 0.0683 & NEPSE autoregressive \\
4  & NEPSE\_Return\_lag3  & 0.0556 & NEPSE autoregressive \\
5  & NEPSE\_Return\_lag4  & 0.0417 & NEPSE autoregressive \\
6  & NEPSE\_Return\_lag1  & 0.0381 & NEPSE autoregressive \\
7  & DayOfWeek            & 0.0294 & Calendar \\
8  & NEPSE\_Vol\_5d       & 0.0218 & NEPSE short-term volatility \\
9  & NEPSE\_Vol\_10d      & 0.0135 & NEPSE volatility regime \\
10 & NEPSE\_Return\_lag5  & 0.0040 & NEPSE autoregressive \\
\bottomrule
\end{tabular}
\end{table}

\begin{figure}[H]
\centering
\includegraphics[width=\textwidth]{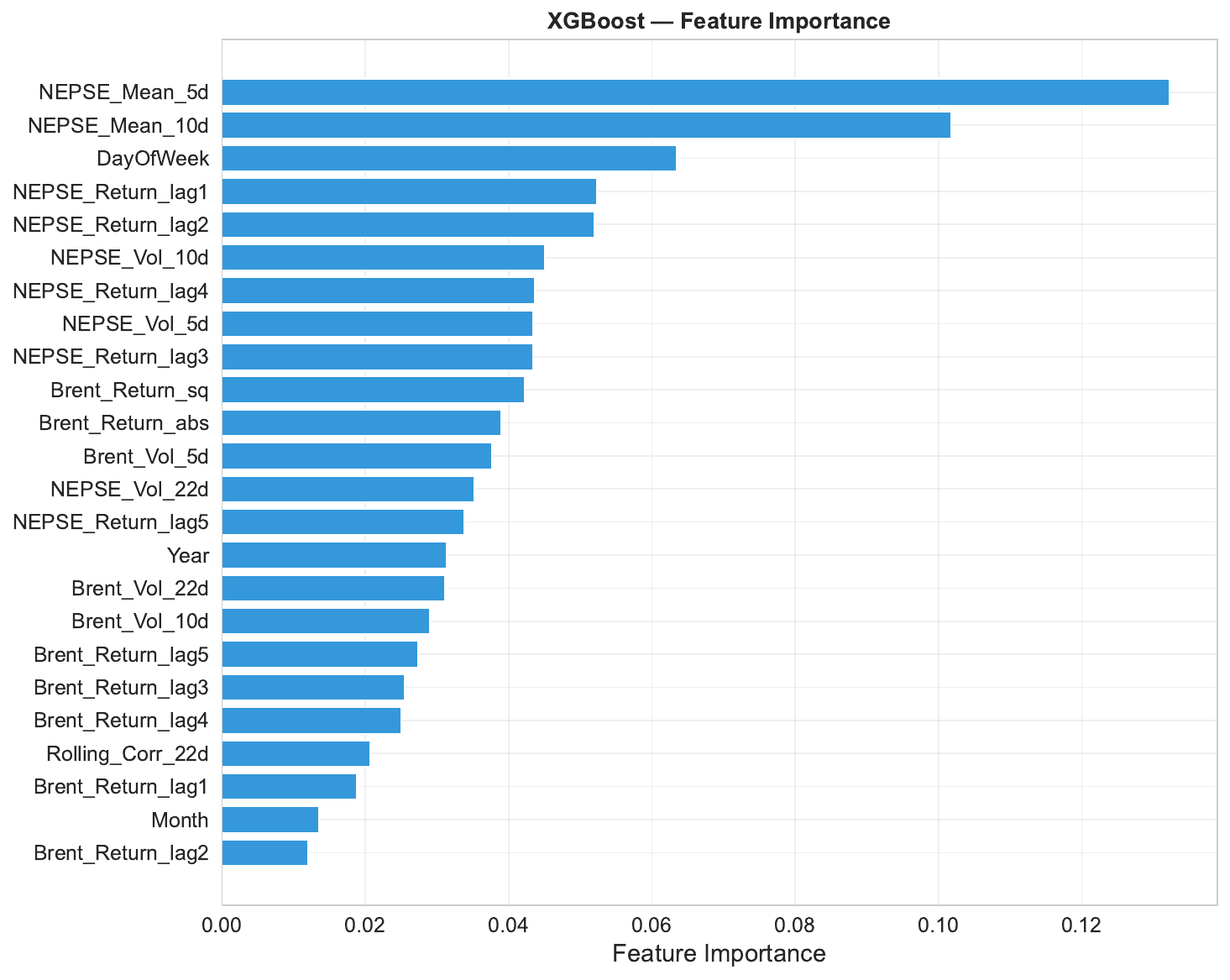}
\caption{XGBoost feature importance ranking.}
\label{fig:xai_fi}
\end{figure}

\begin{figure}[H]
\centering
\includegraphics[width=\textwidth]{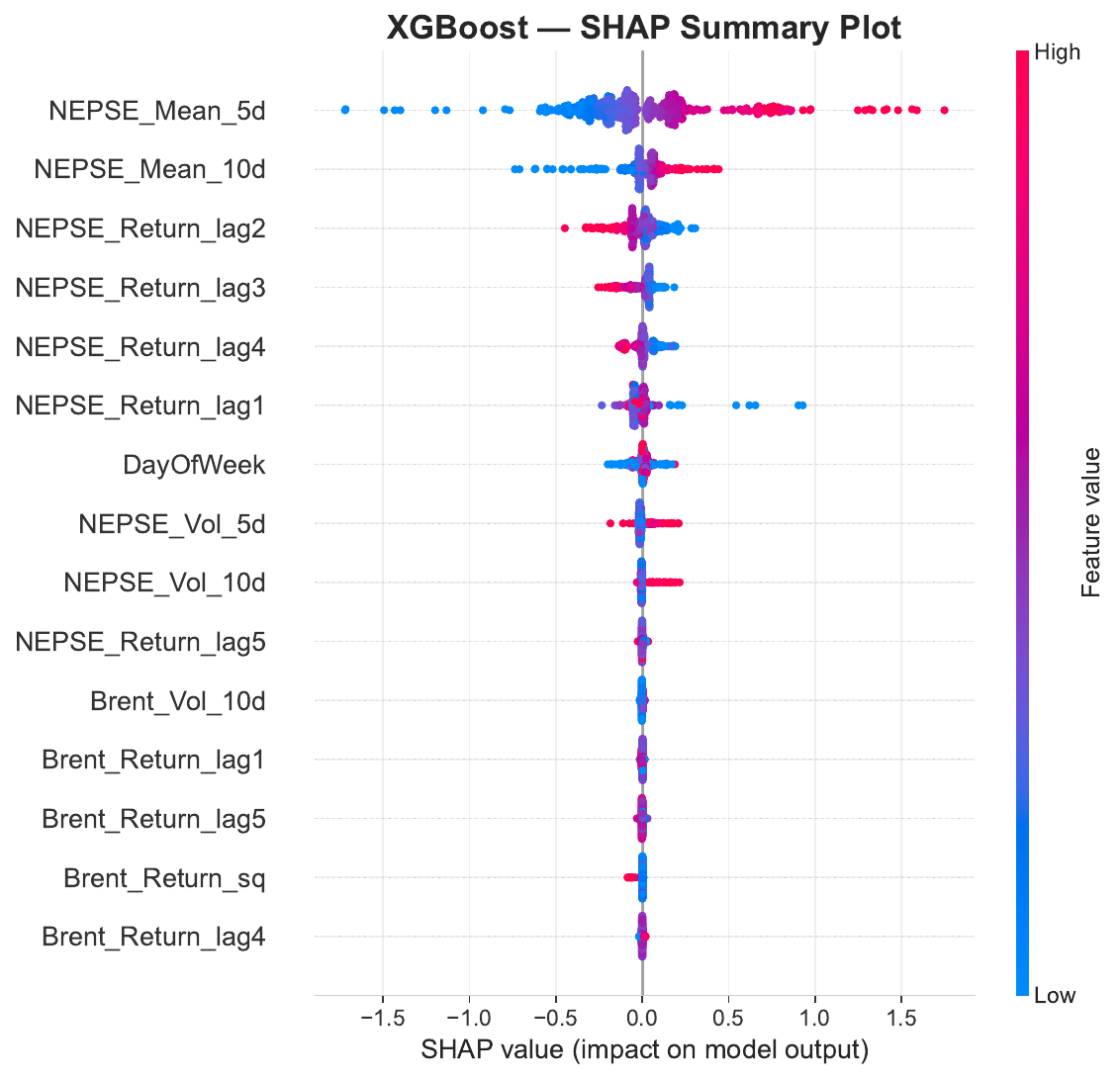}
\caption{SHAP summary plot showing feature importance and directional effects.}
\label{fig:shap_summary}
\end{figure}

\begin{figure}[H]
\centering
\includegraphics[width=0.85\textwidth]{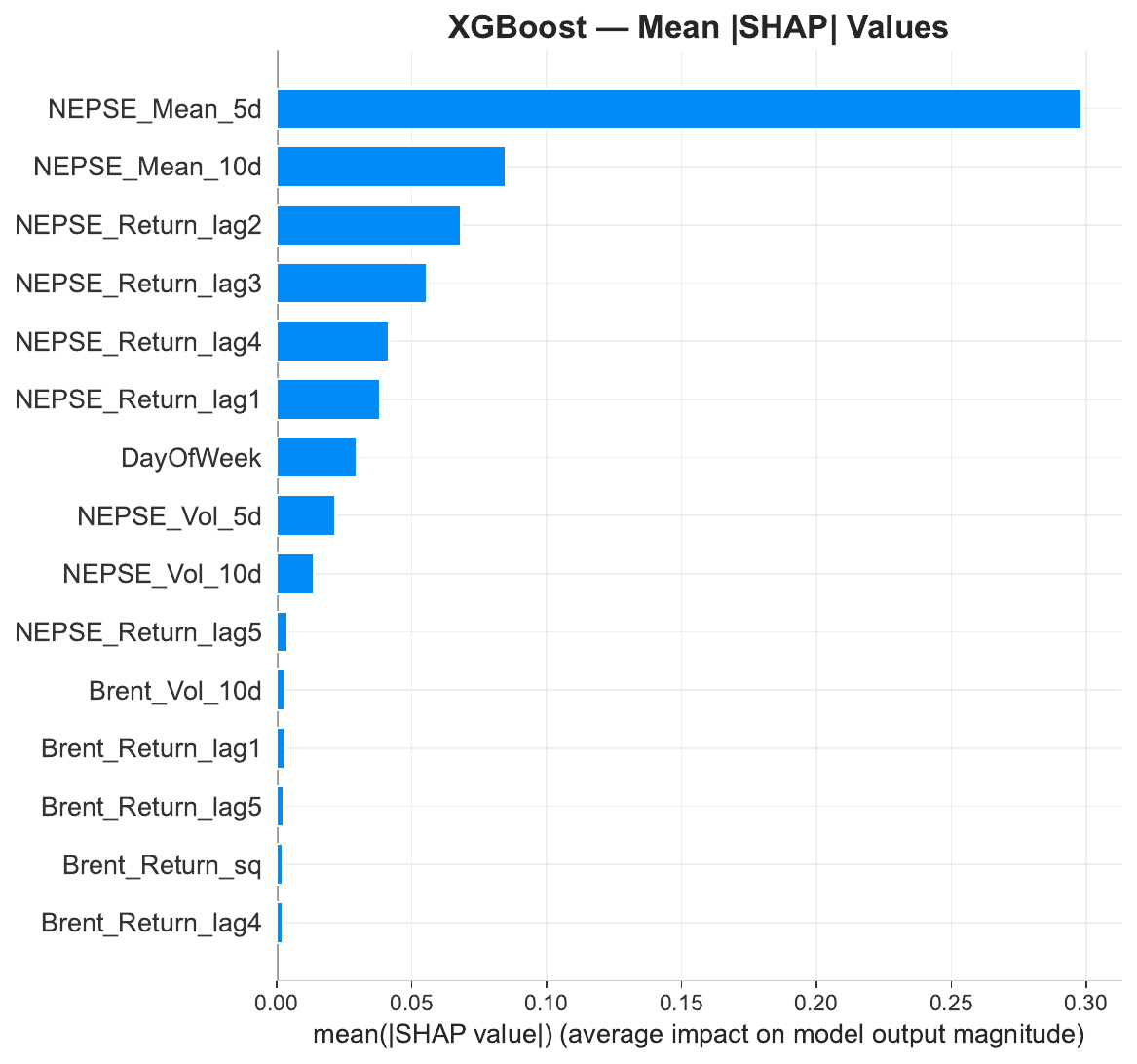}
\caption{SHAP feature importance (mean absolute SHAP values).}
\label{fig:shap_bar}
\end{figure}

\begin{figure}[H]
\centering
\includegraphics[width=\textwidth]{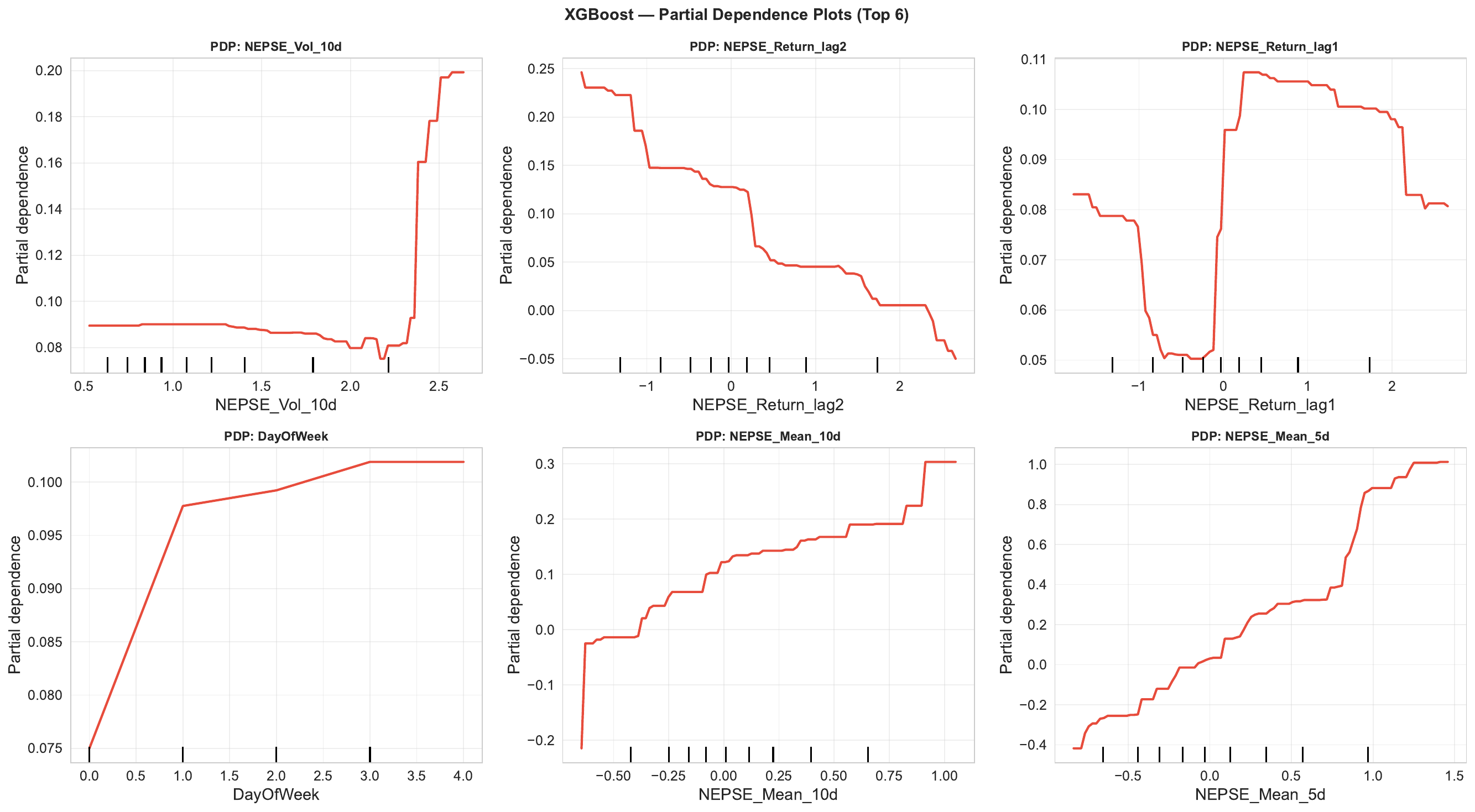}
\caption{Partial Dependence Plots for the top six features.}
\label{fig:pdp}
\end{figure}

\begin{figure}[H]
\centering
\includegraphics[width=\textwidth]{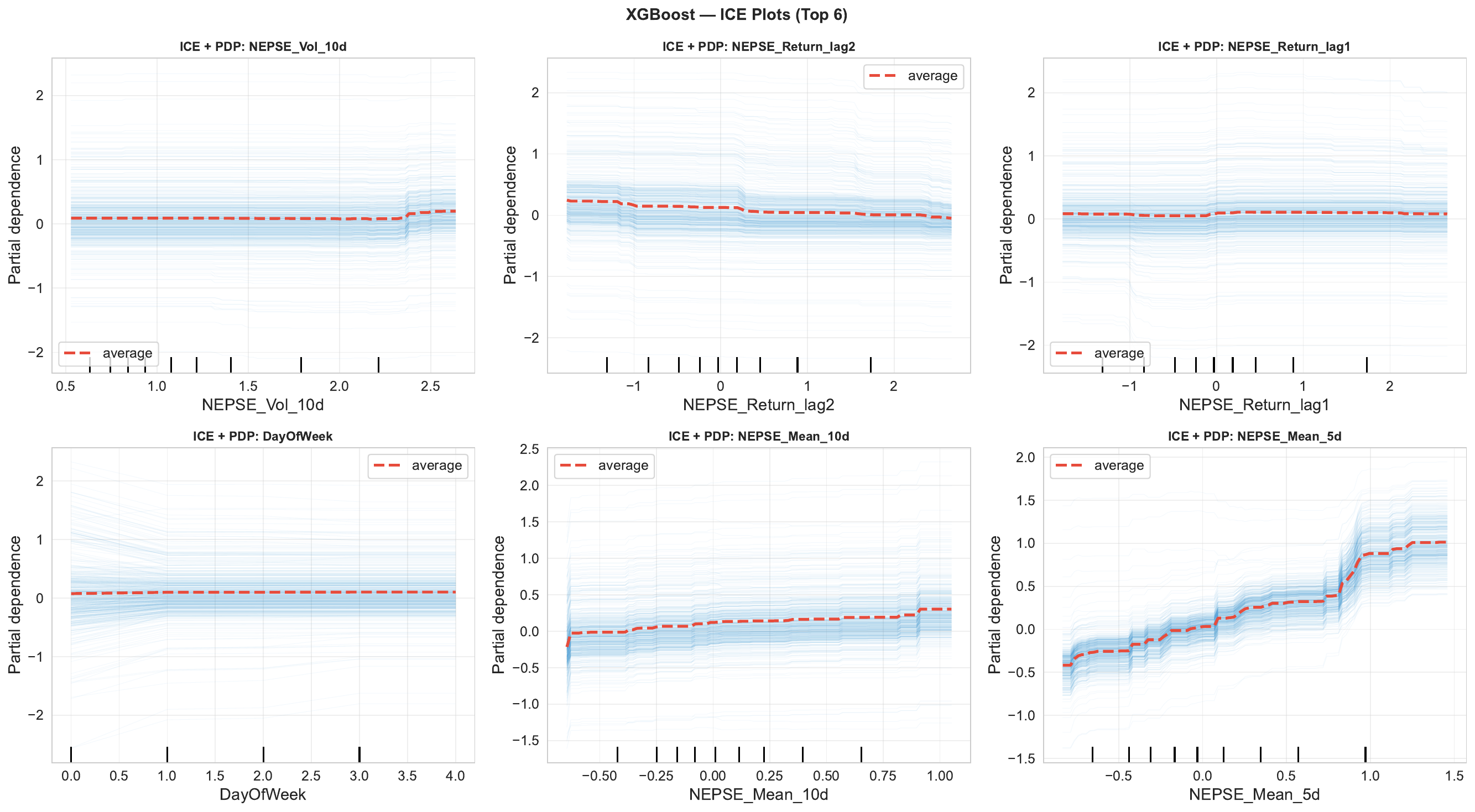}
\caption{Individual Conditional Expectation plots with PDP overlay for the top six features.}
\label{fig:ice}
\end{figure}

\begin{figure}[H]
\centering
\includegraphics[width=\textwidth]{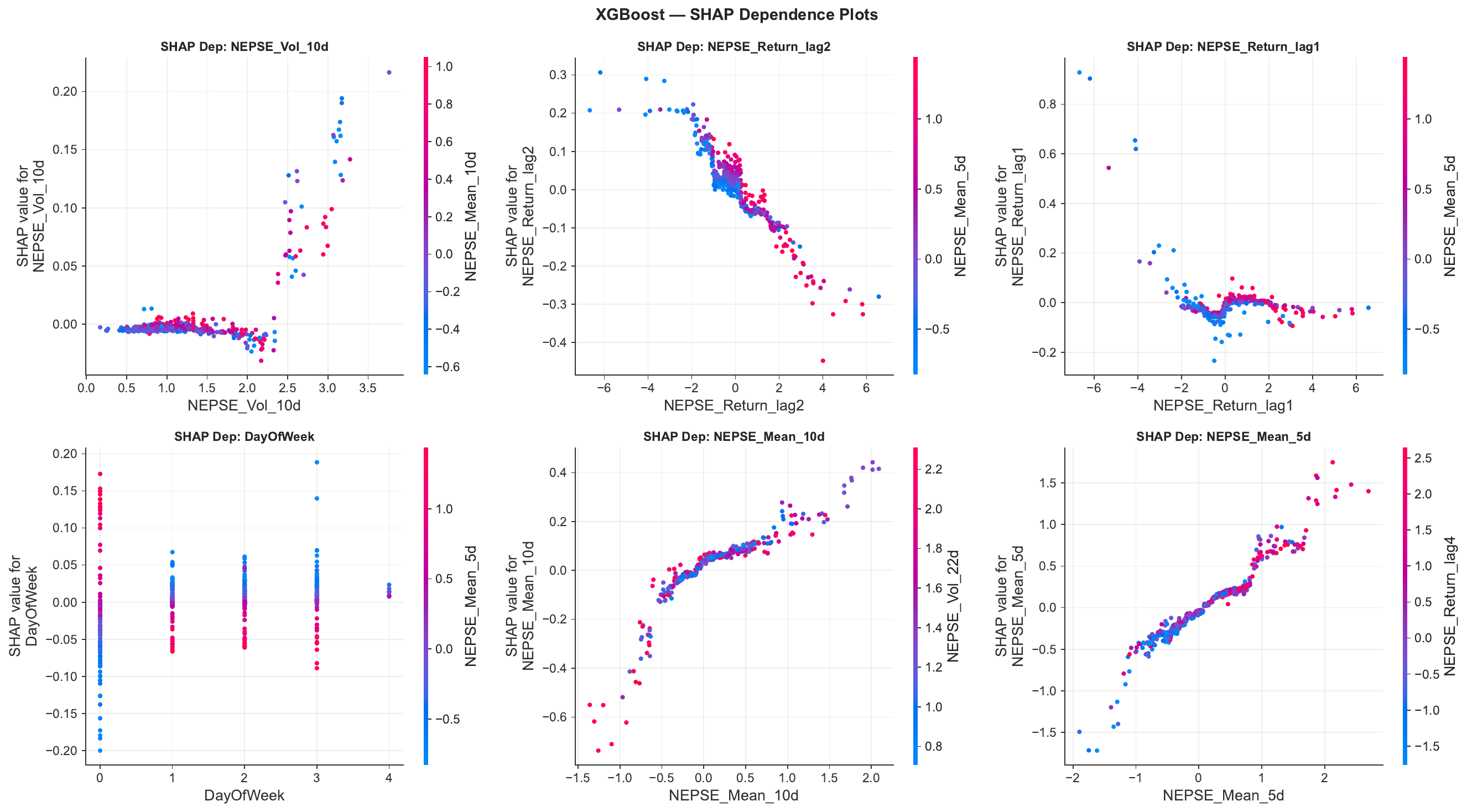}
\caption{SHAP dependence plots for the top six features.}
\label{fig:shap_dep}
\end{figure}

\section{Discussion}

The analysis reveals that there is a statistically significant unidirectional predictive relationship from Brent crude oil prices to NEPSE returns with a nearly four-day transmission lag. It was observed through the Granger causality analysis. Furthermore, the EGARCH model demonstrated that volatility is highly persistent in both markets. The DCC-GARCH model showed that the correlation between the Brent crude oil market and that of the NEPSE is weak. But still, the relationship varies considerably over time. We can conclude from this that external energy market shocks influence Nepal's equity market irregularly, rather than continuously.

Similarly, Machine Learning results have shown that ensemble tree-based algorithms capture nonlinear characteristics in modelling the relationship between Brent crude oil prices and NEPSE returns. Among the models that were evaluated, XGBoost achieved the best predictive performance ($R^2 = 0.29$, DA~=~67\%). Because of this, it was studied using SHAP, Partial Dependence Plots and Individual Conditional Expectation plots. The analysis discovered that NEPSE's own recent momentum and volatility are the strongest predictors for itsfuture returns, and oil-related information is a minor contributor. The importance of lagged NEPSE returns across multiple horizons is consistent with the autoregressive structure identified in the different econometric analyses.

Despite this, there are several limitations that must be acknowledged. The analysis has only considered Brent crude oil prices and the aggregate NEPSE index without incorporating other macro-financial variables that may have an influence on the stock market behavior. Furthermore, the findings are based on historical daily observations. So, it does not account for structural policy changes or future economic shocks. The moderate $R^2$ and directional accuracy obtained from the XGBoost model should be interpreted in context. Although these values exceed the near-zero predictability typical of efficient developed markets, they are consistent with the characteristics of a developing market like NEPSE. In such markets, non-synchronous trading and lower liquidity create temporary statistical pattern which can improve prediction accuracy\cite{lo1990}. The cross-validated results (mean CV $R^2 = 0.28$) confirm that the findings are not driven by a single favourable test split. Future research may extend this framework by including additional variables, sectoral indices, or alternative energy benchmarks and also exploring more advanced explainable artificial intelligence methods parallelly for financial decision support.

\section{Conclusion}

This study was centered around the dynamic relationship that exists between global Brent crude oil prices and the Nepal Stock Exchange (NEPSE). An integrated framework that combined financial econometric models with machine learning techniques was used to analyse daily data over time varying correlations and nonlinear relationships between the two markets.

The findings from the empirical studies suggest that the relationship between Brent crude oil prices and the Nepal Stock Exchange is more sophisticated and complex than suggested by conventional linear analyses. The econometric models demonstrate that Brent crude oil possesses predictive content for NEPSE returns, volatility remains highly persistent, and correlations evolve over time rather than remaining constant. At the same time, the machine learning analysis indicates that the strongest determinants of next-day NEPSE returns are the market's own recent momentum and volatility, whereas oil-related information act as a minor and method-dependent predictor.

The XAI analysis reinforces this interpretation by showing that although Brent crude oil contributes to model predictions, the behaviour of the Nepalese equity market is primarily driven by its own autoregressive structure and short-term market dynamics. As a result, the econometric and explainable machine learning approaches should be viewed as complementing each other's findings rather than competing with each other. While econometric models quantify statistical dependence and volatility transmission, explainable machine learning shows nonlinear interactions and feature contributions that are difficult to capture if we use conventional parametric models only.

Overall, this paper provides empirical evidence that combining financial econometric models with explainable machine learning offers a strong and interpretable framework for the analysis of the dynamic relationship between global commodity markets and equity markets like NEPSE.

\bibliographystyle{plain}

\end{document}